\documentclass[a4paper,11pt,UKenglish]{article}

\usepackage{amssymb,amsmath,natbib,amsthm}
\usepackage{graphicx,ifpdf,color,babel} 
\usepackage[bottom=1in, left=1in, right=1in]{geometry}
\usepackage[figuresright]{rotating}

\newcommand{\pr}{\operatorname{Pr}} 
\newcommand{\mm}[1]{\ensuremath{#1}} 
\newcommand{\CC}{C\nolinebreak\hspace{-.05em}\raisebox{.4ex}{\tiny\bf +}\nolinebreak\hspace{-.10em}\raisebox{.4ex}{\tiny\bf +}}
\begin{document}

\title{Robust Methods for Disease--Genotype
  Association in Genetic Association Studies: Calculate $P$-values
  Using Exact Conditional
  Enumeration instead of Asymptotic Approximations}

\author{M. Langaas and {\O}. Bakke\\
Department of Mathematical Sciences, \\Norwegian University of Science and Technology, \\NO-7491 Norway\\
	   }
\date{\today}
\maketitle

\begin{abstract}
Within the field of genetic association studies, detecting disease--genotype
associations is a primary goal. 
For most diseases, the underlying genetic model is unknown, and
we study seven robust test statistics for monotone disease--genotype association.
For a given test statistic there are many ways to calculate a
$p$-value, but in genetic association studies, $p$-value calculations
have predominantly been based on asymptotic approximations and on
simulated permutation. We show that when the number of permutations tends to infinity, the permutation
$p$-value approaches the exact conditional enumeration $p$-value, and
further that calculating the exact conditional enumeration $p$-value
is much more efficient than performing simulated permutations. 
We then answer two research questions. (i) Which of the seven robust test
statistics under study are the most powerful for monotone genetic models?
(ii)  Based on test size, power, and computational considerations, should
asymptotic approximations or exact conditional enumeration 
be used for calculating $p$-values?
We have studied case--control sample sizes
with 500--5000 cases and 500--15000 controls, and significance levels
from $5\cdot 10^{-8}$ to 0.05, thus our results are applicable to genetic
association studies with only one genetic marker under study,
intermediate follow-up studies, and genome wide association studies. 
Our main findings are as follows. If all monotone genetic models are of
interest, the best performance in the situations
under study is achieved for the robust test statistics based on the
maximum over a range of Cochrane--Armitage trend tests with different 
scores and for the constrained likelihood ratio test.
For significance levels below 0.05, for the
test statistics under study, asymptotic
approximations may give a test size up to 20 times the nominal level,
and should therefore be used with caution. Further, calculating
$p$-values based on exact
conditional enumeration is a powerful, valid and computationally
feasible approach, and we advocate its use in genetic association studies.
\end{abstract}

\section{Introduction}
\label{s:intro}
In genetic association studies the aim is to detect a possible
association between a phenotype and one or many genetic markers. This
can be done for one marker at a time. We will consider biallelic
genetic markers, giving three possible genotypes. For each genetic marker the
following steps can be performed. (i) First an hypothesis test
situation is specified. (ii) This guides the choice of a test
statistic. (iii) Then a method to calculate a
$p$-value is chosen. (iv) Finally, the calculated $p$-value is compared to a chosen significance
level to arrive at a conclusion.

A hypothesis test situation may be formulated as ``no association between
the disease and the genetic marker'' versus ``association
between the disease and the genetic marker''.  In some genetic studies,
in particular for monogenic diseases, the effect of the disease allele
on the disease phenotype may be known to follow a specific
\emph{genetic model}, such as dominant, recessive,
additive, or multiplicative \citep[e.g.][]{Camp1997}. The genetic model can be used
as alternative hypothesis to construct a test statistic
tailored to detect this type of disease--genotype association.
For genome-wide association (GWA) studies the underlying genetic model
of a disease allele is seldom known \citep[e.g.][]{DevlinRoeder1999}. Therefore, it is desirable to
base the statistical inference on a test statistic that give high power over a wide range of genetic
models. These types of test statistics are commonly referred to as
\emph{robust} test statistics \citep{FreidlinZhengLiGastwirth2002}. We will restrict our attention to
performance under what we will call \emph{monotone} genetic models,
under which the genetic effect of the heterozygous genotype lies
between the two homozygous genotypes. This means that we are not considering
overdominant models. The first research questions we want to
address in this presentation is which of the available popular robust test statistics that are
most suitable for use when the underlying genetic model is unknown,
but monotone.

Two approximations that have been used to calculate $p$-values for
disease--genotype association are asymptotic methods and simulated
permutations \citep{Sladek2007}. We will argue that exact conditional
enumeration yields the same $p$-values as permutation when the number
of permutations tends to infinity, and is also less computationally
intensive than simulated permutations. We will therefore consider asymptotic
methods and exact conditional enumeration. Many other ways of
calculating $p$-values exists, see \cite{LangaasBakkeMAX3} for a
presentation of unconditional enumeration methods for
discrete distributions suitable for small to moderate sample sizes.

Exact conditional enumeration is a general method by which the $p$-value of an
outcome is calculated based on a test statistic and the conditional
probabilities of all possible outcomes of the conditional experiment in question.
The most popular use of exact conditional
enumeration is the Fisher exact test. There is a large body of literature on hypothesis testing in $2\times
2$ contingency tables, where conditional tests are often found to be
less powerful than unconditional alternatives, as described by
\cite{MehrotraChanBerger2003} and \cite{Lydersenetal2009}. 
Our focus is on disease--genotype association for a biallelic marker
in a case--control setting, which means that the data can be presented in
a $2\times3$ contingency table. The conditioning is done on the column
margins, and the
conditional probability of an outcome is a trivariate
hypergeometric probability. Due to the less discrete nature of higher
order contingency tables, some researchers have found
that the power disadvantage of the
conditional test tends to be less pronounced than for the $2 \times 2$
contingency table \citep{MehtaHilton1993}.
For larger contingency tables exact conditional enumeration is believed to
require substantial computer resources (enumeration and summation of
probabilities) and thus not to be feasible for
testing for disease--genotype association. We will show that even for
large sample sizes, 5000 cases and at least 5000 controls in a
$2\times 3$ contingency table, exact
conditional enumeration can be performed in a fraction of a second on
a standard computer. Turning to the asymptotic methods, it is known that asymptotic
methods will not preserve test size in situations where the asymptotic
approximation is poor. The second research question we want to address
is therefore whether
asymptotic or exact conditional enumeration methods are most suitable
to use in terms of power, test size and computational resources
when testing disease--genotype association with a robust test
statistic.

In GWA studies multiple testing correction is commonly
performed by controlling the familywise error rate by the Bonferroni 
method. When a single genetic marker is studied, a significance
level of $\alpha=0.05$ is commonly used. For larger
candidate studies, or follow-up studies, with 10--1000 genetic markers
under study, significance levels of the order $5\cdot 10^{-3}-5\cdot
10^{-5}$ may be used. For genome-wide association (GWA) studies
with ten thousand to one million genetics markers, significance levels $5 \cdot 10^{-6}-5 \cdot
10^{-8}$ have been used \citep{WTCCC2007}. \cite{Dudbridge2008} advocated using significance level
$7.2\cdot 10^{-8}$ for general GWA studies. When we investigate our
two research questions we will study significance levels in the 
range $5\cdot 10^{-2}- 5 \cdot 10^{-8}$, and sample sizes in the range
500--5000 cases and 500--15000 controls.

The presentation is organized as follows. In Section
\ref{sec:methods} we present notation, data structure, test
statistics, methods for calculating $p$-values and estimation of
power. We then present the results of a large simulation study to compare the power of different
robust test statistics combined with either the asymptotic method or the exact conditional
enumeration method in Section \ref{sec:power}. We discuss in Section
\ref{sec:discuss} and conclude in Section \ref{sec:conclude}.

\section{Methods}\label{sec:methods}
\subsection{Notation and data}
We will to some extent adopt the notation of
\cite{JooRobustWellcome2009} and \cite{LangaasBakkeMAX3}.
We assume that genotypes and a dichotomous phenotype (disease)
are collected in a case--control study, and that the genotype data are from biallelic
genetic markers with alleles $a$ and $A$. 
For each genetic marker, we assume that $A$ is the high risk allele,
and index the three genotypes $aa$, $aA$, and $AA$, by 0, 1, 2,
respectively. Further, for each genetic marker let $g_0$, $g_1$ and $g_2$ be the
genotype frequencies for the three genotypes in the population under study, $p_0$,
$p_1$, and $p_2$ the genotype frequencies of the case population
(disease phenotype) and $q_0$, $q_1$, and $q_2$ the genotype
frequencies of the control population. 

For each genetic marker we may present the collected data in a 
$2\times 3$ contingency table (Table~\ref{tab:2x3notation}). The number of cases and controls with genotype~$i$ is denoted by $x_i$ and $y_i$, respectively, and the total number of cases and controls with genotype~$i$ by $m_i=x_i+y_i$, \ $i=0$, 1, 2.  Let $n_1=x_0+x_1+x_2$ denote the total number of cases, $n_2=y_0+y_1+y_2$ the total number of controls, and let $N=n_1+n_2=m_0+m_1+m_2$.
\begin{table}
\caption{Notation for $2\times 3$ table, case--control data.} 
\begin{center}
\begin{tabular}{lcccc}\hline
      & \multicolumn{3}{c}{Genotype}&\\
\cline{2-4}
      & $aa$ & $aA$ &$AA$&Total\\
      \hline
      Case & $x_0$&$x_1$&$x_2$&$n_1$\\
      Control & $y_0$&$y_1$&$y_2$&$n_2$\\
      \hline
      Total & $m_0$&$m_1$&$m_2$&$N$\\ \hline
    \end{tabular}
\end{center}
\label{tab:2x3notation}
\end{table}

The vectors $(x_0,x_1,x_2)$ and $(y_0,y_1,y_2)$ are considered to be
independent and trinomially distributed with parameters
$(n_1;p_0,p_1,p_2)$ and $(n_2;q_0,q_1,q_2)$, respectively. The
probability of a disease--genotype table with entries
$z=(x_0,x_1,x_2,y_0,y_1,y_2)$ for a given set of parameters
$\theta=(p_0,p_1,p_2,q_0,q_1,q_2)$ is
\begin{equation}
 \pr_\theta(Z=z)=\frac{n_1!}{x_0!x_1!x_2!}p_0^{x_0}p_1^{x_1}p_2^{x_2}
  \cdot\frac{n_2!}{y_0!y_1!y_2!}q_0^{y_0}q_1^{y_1}q_2^{y_2}.\label{eq:probZ}
\end{equation}

\subsection{Statistical Hypotheses and Genetic Models}\label{sec:genmod}
Let $k$ be the disease prevalence, and $f_i$ be the penetrance, i.e. the conditional
probability of disease given genotype $i$.
Then $p_ik=f_ig_i$ and $q_i(1-k)=(1-f_i)g_i$ for $i=0$, 1, 2.
The null hypothesis that we will investigate is
that the penetrances are equal for the three possible genotypes, 
\begin{equation}
 f_0=f_1=f_2, \label{eq:H0}
\end{equation}
which can be shown to be equivalent to $p_i=q_i$ for $i=0$, 1, 2.
Further, denote by $\lambda_1=f_1/f_0$ and
$\lambda_2=f_2/f_0$ the genotype relative risks.
We define a {\em monotone genetic model} to satisfy
\begin{equation}
f_0\le f_1 \le f_2,\label{eq:H1}
\end{equation}
or  alternatively $1 \le \lambda_1 \le
\lambda_2$, which can be shown to be equivalent to $p_0/q_0\leq p_1/q_1\leq p_2/q_2$. As alternative hypotheses we consider monotone genetic models where at least one of the equalities are strict. 
Those models can be parameterized by
\begin{equation}
\lambda_1=1-\delta+\delta\lambda_2, \label{eq:obmod}
\end{equation}
where $\lambda_2>1$ and $0\leq\delta\leq1$. The value $\delta=0$ yields the recessive
genetic model, $\delta=1/2$ the additive genetic model, and $\delta=1$ the
dominant genetic model. We will refer to the genetic model with
$\delta=1/4$ as semi-recessive and $\delta=3/4$ as semi-dominant.

\subsection{Test Statistics and Asymptotic Distributions}\label{statistics}
We now consider test statistics for testing the null hypothesis \eqref{eq:H0} against the alternative of a general monotone genetic model~\eqref{eq:obmod} or some specified monotone genetic model.
The potential high risk allele is often unknown. Therefore all tests will be two-sided, in the sense that the conclusion of the test will be the same if the data for each homozygote are swapped.
 
\subsubsection{Cochran--Armitage Trend Test {(CATT)}}
The Cochran--Armitage test for trend (CATT) \citep{Armitage1955,Cochran1954,sasieni1997genotypes,slager2001case} is often used to test the null hypothesis \eqref{eq:H0} against one of the common (recessive, additive, dominant) genetic models (alternative hypotheses) in \eqref{eq:obmod}. 
It is based on the statistic $\sum_{i=0}^2s_i(x_i/n_1-y_i/n_2)$, where $s_0$, $s_1$, $s_2$ are scores appropriate for the alternative hypothesis in question. Standardizing and replacing unknown parameters $p_i$, $q_i$ by estimators $m_i/N$, we obtain the CATT test statistic,
\[
 \text{CATT}=\frac{\sum_{i=0}^2s_i(n_2x_i-n_1y_i)}
  {\sqrt{n_1n_2\Big(\sum_{i=0}^2s_i^2m_i-\frac1N\big(\sum_{i=0}^2s_im_i\big)^2\Big)}},
\]
which asymptotically has a standard normal distribution under the null
hypothesis. The absolute value of $\text{CATT}$ is invariant to linear
transformations of the scores, so they are chosen
$(s_0,s_1,s_2)=(0,s,1)$, and we use the notation $\text{CATT}_s$ for CATT with those scores. The
value of $s$ is chosen as $s=0$, $1/2$, 1 for the recessive,
additive and dominant model of \eqref{eq:obmod}, respectively
\citep{zheng2003choice}. The index $s$ thus denotes which genetic
model (alternative hypothesis) is used. A large value of
$|\text{CATT}_s|$ indicates rejection of the null hypothesis.


We will study the $\text{CATT}_{1/2}$ test statistics further.

\subsubsection{Pearson Chi-Squared Test {(Pearson)}}\label{pearson}
The well-known Pearson chi-squared test statistic 
\[
 \sum_{i=0}^2\bigg(\frac{(x_i-m_in_1/N)^2}{m_in_1/N}+\frac{(y_i-m_in_2/N)^2}{m_in_2/N}\bigg)
\]
is not tailored to be powerful for monotone genetic models~\eqref{eq:H1} in particular but rather to test against the more general alternative that $f_0$, $f_1$ and $f_2$ are not all equal.
Under the null hypothesis, the Pearson test statistic for our $2\times 3$ situation asymptotically follows a chi-squared distribution with two degrees of freedom. A large value indicates rejection of the null hypothesis.

\subsubsection{Minimum $p$-Value Test {(MIN2)}}
The statistic MIN2 is defined as the minimum of the asymptotic $p$-values of $\text{CATT}_{1/2}$ and of the Pearson chi-squared statistic. It is not a valid $p$-value itself, but its asymptotic distribution under the null hypothesis is given by
\[
 \pr(\text{MIN2}\leq t)\to\frac12t+\frac12e^{-q/2}-\frac1{2\pi}\int_q^{-2\ln t}e^{-v/2}\arcsin\Big(\frac{2q}v-1\Big)dv,
\]
where $q$ is the $1-t$ quantile of the chi-squared distribution with one degree of freedom \citep{JooRobustWellcome2009}. A small value of MIN2 indicates rejection of the null hypothesis.

\subsubsection{Maximum test {(MAX3)}}
The statistic
$\text{MAX3}=\max(|\text{CATT}_0|,|\text{CATT}_{1/2}|,|\text{CATT}_1|)$
was proposed as an alternative to CATT when the genetic model is
unknown but monotone, with emphasis on the recessive, additive or dominant
model \citep{FreidlinZhengLiGastwirth2002}. Asymptotically, $\text{CATT}_{1/2}$ is a linear combination of $\text{CATT}_0$ and $\text{CATT}_1$, and $(\text{CATT}_0,\text{CATT}_1)$ has a bivariate normal asymptotic distribution \citep{Rassoc2010}. The asymptotic $p$-value of MAX3 can be found as the probability of a bivariate normal pair lying outside a region, which is in general hexagonal, in the plane. Specifically,
\begin{multline*}
 \pr(\text{MAX3}\geq t)
  \to1-2\int_0^{(1-\omega_1)t/\omega_0}
   \phi(x)\Bigg(\Phi\bigg(\frac{t-\rho x}{\sqrt{1-\rho^2}}\bigg)
    -\Phi\bigg(\frac{-t-\rho x}{\sqrt{1-\rho^2}}\bigg)\Bigg)dx\\
   -2\int_{(1-\omega_1)t/\omega_0}^t
    \phi(x)\Bigg(\Phi\bigg(\frac{(t-\omega_0x)/\omega_1-\rho x}{\sqrt{1-\rho^2}}\bigg)
     -\Phi\bigg(\frac{-t-\rho x}{\sqrt{1-\rho^2}}\bigg)\Bigg)dx,
\end{multline*}
where
\[
 \omega_0=\sqrt{\frac{g_2(1-g_2)}{g_0(1-g_0)+g_2(1-g_2)+2g_0g_2}}\qquad\text{and}\qquad
 \omega_1=\sqrt{\frac{g_0(1-g_0)}{g_0(1-g_0)+g_2(1-g_2)+2g_0g_2}}
\]
are the coefficients making $\text{CATT}_{1/2}\to\omega_0\text{CATT}_0+\omega_1\text{CATT}_1$ asymptotically,
\begin{equation}
 \rho=\sqrt{\frac{g_0g_2}{(1-g_0)(1-g_2)}}\label{eq:rho}
\end{equation}
is the asymptotic correlation coefficient of $\text{CATT}_0$ and
$\text{CATT}_1$ under the null hypothesis \eqref{eq:H0}, and $\phi$ and $\Phi$ are the standard normal pdf and cdf, respectively \citep{Rassoc2010}. When the asymptotic $p$-value is computed, $g_0$, $g_1$ and $g_2$ must be replaced by their consistent estimators $m_0/N$, $m_1/N$ and $m_2/N$, respectively.

\subsubsection{Constrained Maximum Test {(CMAX)}}\label{cmax}
The Pearson chi-squared test statistic (Section~\ref{pearson}) is equal to $\text{CATT}_s^2$, where the score is determined by the data, $s=(x_1/m_1-x_0/m_0)/\allowbreak{(x_2/m_2-x_0/m_0)}$ \citep{zheng2009pearson}, which can be viewed as an estimator for $(f_1-f_0)/(f_2-f_0)$. Also, this is the $s$ maximizing $\text{CATT}_s$ \citep{zheng2009pearson}. This motivates a test statistic,
\[
 \text{CMAX}=\max_{0\leq s\leq1}\text{CATT}_s=
  \begin{cases}\text{Pearson}&\text{if $0<s<1$}\\
   \max(\text{CATT}_0^2,\text{CATT}_1^2)&\text{otherwise}\end{cases}
\]
\citep[the second equality was shown by][]{zheng2009pearson}.

The distribution of CMAX can be seen as a mixture of the distribution
of the Pearson statistic and the distribution of
$\max(\text{CATT}_0^2,\text{CATT}_1^2)$, with weights $w'=\pr(0<s<1)$
and $1-w'$, respectively. Applying a two-dimensional version of the
central limit theorem to the trinomial vector $(x_0,x_1,x_2)$, the
asymptotic value of $w'$ is found as the integral of a binormal
density over the region $0<s<1$ in the $(x_0,x_2)$ plane, giving
$w'\to w=\frac1\pi\arccos\rho$ (see~\eqref{eq:rho}).
Under the null hypothesis, the Pearson statistic has the chi-squared
distribution with two degrees of freedom, and the asymptotic
distribution of $(\text{CATT}_0,\text{CATT}_1)$ is standard binormal
with correlation $\rho$. The asymptotic $p$-value of CMAX is
\[\pr(\text{CMAX}\geq t)=1-wF(t)-2(1-w)\int_0^{\sqrt t}\phi(x)\Bigg(\Phi\bigg(\frac{\sqrt t-\rho x}{\sqrt{1-\rho^2}}\bigg)-\bigg(\frac{-\sqrt t-\rho x}{\sqrt{1-\rho^2}}\bigg)\Bigg)dx,
\]
where $\phi$ and $\Phi$ are the standard normal pdf and cdf, respectively, and $F$ the cdf of the chi-squared distribution with two degrees of freedom. Again, $g_0$, $g_1$ and $g_2$ enter into $\rho$, and must be replaced by their consistent estimators $m_0/N$, $m_1/N$ and $m_2/N$, respectively, to compute an asymptotic $p$-value.

\subsubsection{Constrained Likelihood Ratio Test {(CLRT)}}
CLRT is the likelihood ratio test statistic, but the maximum under the alternative hypothesis is taken under the constraint of the genetic model being monotone (see Section~\ref{sec:genmod}). The log likelihood given by the data from the two independent trinomial distributions is
\[
 l=\text{constant}+\sum_{i=0}^2x_i\ln p_i+\sum_{i=0}^2y_i\ln q_i.
\]
It is maximized by $p_i=x_i/n_1$ and $q_i=y_i/n_2$, giving a maximum (omitting the constant)
\[
 l_1=x_0\ln x_0+x_1\ln x_1+x_2\ln x_2+y_0\ln y_0+y_1\ln y_1+y_2\ln y_2-n_1\ln n_1-n_2\ln n_2.
\]
Under the constraint of a recessive model, $f_0=f_1$, or equivalently, $p_0/q_0=p_1/q_1$, the maximum is obtained at
\[
 (p_i,q_i)=\frac{m_i}{m_0+m_1}\left(\frac{x_0+x_1}{n_1},\frac{y_0+y_1}{n_2}\right),
  \quad\text{$i=0$, 1}\qquad\text{and}\qquad p_2=\frac{x_2}{n_1},\quad q_2=\frac{y_2}{n_2},
\]
which gives the maximum
\begin{multline*}
 l_\text{rec}=(x_0+x_1)\ln(x_0+x_1)+x_2\ln x_2+(y_0+y_1)\ln(y_0+y_1)+y_2\ln y_2\\
  +m_0\ln m_0+m_1\ln m_1-(m_0+m_1)\ln(m_0+m_1)-n_1\ln n_1-n_2\ln n_2.
\end{multline*}
Similarly, under the constraint of a dominant model, $f_1=f_2$, or equivalently, $p_1/q_1=p_2/q_2$, the maximum is obtained at
\[
  p_0=\frac{x_0}{n_1},\quad q_0=\frac{y_0}{n_2}\qquad\text{and}\qquad
 (p_i,q_i)=\frac{m_i}{m_1+m_2}\left(\frac{x_1+x_2}{n_1},\frac{y_1+y_2}{n_2}\right),\quad\text{$i=1$, 2},
\]
which gives the maximum
\begin{multline*}
 l_\text{dom}=x_0\ln x_0+(x_1+x_2)\ln(x_1+x_2)+y_0\ln y_0+(y_1+y_2)\ln(y_1+y_2)\\
  +m_1\ln m_1+m_2\ln m_2-(m_1+m_2)\ln(m_1+m_2)-n_1\ln n_1-n_2\ln n_2.
\end{multline*}
Under the null hypothesis~\eqref{eq:H0}, the maximum is obtained at $p_i=q_i=m_i/N$, giving the maximum
\[
 l_0=m_0\ln m_0+m_1\ln m_1+m_2\ln m_2-N\ln N.
\]
Then
\[
 \text{CLRT}=\begin{cases}-2(l_1-l_0)&\text{if $0\leq s\leq 1$}\\
  -2(\max(l_\text{rec},l_\text{dom})-l_0)&\text{otherwise,}\end{cases}
\]
where $s$ is the data-driven score defined in Section~\ref{cmax}. This
is the same statistic as obtained by \cite{wang2005constrained}, who
showed that CLRT has the same asymptotic distribution under the null hypothesis as described for CMAX (Section~\ref{cmax}).

\subsubsection{Maximin Efficiency Robust Test {(MERT)}}
A maximin efficiency robust test \citep{gastwirth1985use} can be constructed from $\text{CATT}_0$ and $\text{CATT}_1$, giving $\text{MERT}=(\text{CATT}_0+\text{CATT}_1)/\sqrt{2(1+\rho)}$, where $\rho$ is defined in~\eqref{eq:rho} \citep{zheng2006comparison}. It has an asymptotic standard normal distribution under the null hypothesis. A large value of $|\text{MERT}|$ indicates rejection of the null hypothesis.

\subsection{Using Conditional Enumeration to Calculate $p$-Values}\label{pcond}
When an outcome $z=(x_0,x_1,x_2,y_0,y_1,y_2)$, is presented as a contingency table (Table \ref{tab:2x3notation}) the column margins are $m_0=x_0+y_0$, $m_1=x_1+y_1$ and $m_2=x_2+y_2$. When we condition on the column margins $M(z)=(m_0,m_1,m_2)$, the probability under the null hypothesis of an outcome $z$ is a trivariate hypergeometric probability
\begin{equation}
 \pr(Z=z\mid M(Z)=M(z))= \frac{\binom{m_0}{x_0}\binom{m_1}{x_1}\binom{m_2}{x_2}}{\binom{N}{n_1}},\label{eq:c}
\end{equation}
showing that the column margins are sufficient statistics for the genotype frequencies, which would otherwise be nuisance parameters.
Any test statistic $T$ (with, say, large values indicating rejection of the null hypothesis) defines a $p$-value of an outcome $z$ conditioned on its column margins $M(z)$. It can be calculated by the sum
\begin{equation}
 p(z)= \pr(T(Z)\ge T(z)\mid M(Z)=M(z))=\sum_{T(z')\geq T(z)}\pr(Z=z'\mid M(z')=M(z)). \label{eq:Cpvalue}
\end{equation}
 The number of summands in~\eqref{eq:Cpvalue} is much smaller
 than it would have been without conditioning, making summation also
 feasible for relatively large studies. \cite{Bakke2012} found a
 formula for the maximum number of summands in~\eqref{eq:Cpvalue}. For unbalanced sample sizes where $n_2\ge 2n_1$ the
 maximum number of summands is simply $\binom{n_1+2}{2}$. Numerical examples are presented in Table
 \ref{tab:ctabs}. 

We have seen that the outcome of an experiment can be presented as a contingency table $z=(x_0,x_1,x_2,y_0,y_1,y_2)$. The outcome may alternatively be given on the individual level as two vectors of length $N$, one giving the disease status and one giving genotype status. Thus, entry $l$ in the disease vector gives the disease status of individual $l$ and entry $l$ in the genotype vector gives the coded genotype of individual $l$. In permutation testing we generate $b$ new outcomes of our experiment by permuting (shuffling) the genotypes vector, while keeping the disease vector fixed. This gives $b$ new contingency tables with the same margins as the observed contingency table.
The permutation $p$-value is given as the proportion of the $b+1$
outcomes (the original outcome and the $b$ permutation outcomes)
having a value of the test statistic $T$ greater than or equal to that
of the original outcome. The permutation $p$-value is valid
\citep{PhipsonSmyth2010}. When $b$ tends to infinity the permutation
$p$-value equals the exact conditional enumeration $p$-value. This can
be seen by the fact that the permutation procedure is a trivariate
hypergeometric experiment, drawing genotypes of the $n_1$ cases from
the $m_0$, $m_1$, $m_2$ of each genotype. \cite{Moldovan2013} show in a worked-through example how to calculate the exact conditional enumeration $p$-value using the MAX3 test statistic.

We recommend using the exact conditional enumeration $p$-value, and
not the simulated permutation $p$-value, based on the following arguments. If
the permutation algorithm is run more than once for the same observed
outcome, this may result in a different simulated permutation $p$-value for each run, which for a given significance level may lead to different hypothesis testing decisions.
For GWA studies a significance level of $5\cdot 10^{-8}$ is routinely
used. To be able to arrive at a $p$-value below this significance
level $b$ must at least be $2\cdot 10^7$. Using permutation with very
large values of $b$ is very inefficient compared to using~\eqref{eq:Cpvalue} directly, as can also be seen from Table~\ref{tab:ctabs}.

\begin{table}
\caption{Maximum number $N^*$ of tables with given column
  margins for sample size $n_1$ cases and $n_2$ controls. This will be the maximum number of summands when
  calculating the exact conditional enumeration $p$-value in Equation~\eqref{eq:Cpvalue}. The notation $\geq i$ means that $N^*$ is the
  same for all $n_2\geq i$. }
\begin{center}
\begin{tabular}{rrrrrrrrrrrrrr}
  \hline
$n_1$& 500 & 500 &  1000 & 1000  & 5000  & 5000 \\ 
$n_2$ & 500 & $\ge$1000 & 1000 & $\ge$2000 & 5000 & $\ge$10000  \\ 
$N^*$ & 83834 & 125751  & 334334 & 501501  & 8338334 & 12507501 \\ 
 \hline
\end{tabular}
\end{center}
\label{tab:ctabs}
\end{table}

\subsection{Validity of $p$-values}\label{sec:valid}
For a chosen significance level $\alpha$ and an outcome $z$, the null
hypothesis is rejected if the $p$-value $p(z)\leq\alpha$. For a test
to keep its size, the probability of rejection under the null
hypothesis should be less than or equal to $\alpha$,
i.e. $\pr(p(Z)\leq\alpha)\leq\alpha$ for all $\alpha$ and all parameters under the null hypothesis. Such a
$p$-value is called \emph{valid}
\citep[][p. 397]{casella2001statistical}.

When calculating a $p$-value based on the asymptotic distribution of a
test statistic, there is no reason to believe that this will be a valid
$p$-value for the sample size under study. The conditional $p$-value defined in Section~\ref{pcond}, is on the other hand always valid, not only considered as a $p$-value when the experiment is conditioned on $M(Z)$, but also considered as a $p$-value for the original experiment (here case--control) \cite[][p. 399]{casella2001statistical}.

\subsection{Power}
 Desirable properties of a $p$-value are validity and high power (the
 probability to reject $H_0$). If the sample space is discrete, the
 power at a parameter vector $\theta$ of a test defined by $p(Z)$ for a given $\alpha$ is
\begin{equation}
 \gamma(\mm{\theta})=\pr_\theta(p(Z)\leq\alpha)=\sum_{p(z)\leq\alpha}\pr_\theta(\mm{Z}=\mm{z}), \label{eq:exactpower}
\end{equation}
where the probabilities depend on the parameter vector $\mm{\theta}$, and the summation is
over all outcomes $z$ having a $p$-value not greater than
$\alpha$. The test size is $\sup_{\theta\in\Theta_0}\gamma(\theta)$,
where the supremum is taken over all parameter vectors under the null hypothesis.

In our set-up, explained in detail in Section~\ref{sec:power}, the
number of summands in~\eqref{eq:exactpower} is maximally
$\binom{n_1+2}{n_1}\binom{n_2+2}{n_2}$, which becomes too large for
practical use for the sample sizes we consider.
We instead estimate test size and power using
simulation. We base our calculations on $b$ independent random draws from the probability 
distribution for the data~\eqref{eq:probZ}. Let $p(z_i)$ be the
calculated $p$-value for drawing $i$. Then the estimated power,
$\widehat{\gamma(\mm{\theta})}$, is
\begin{equation}
\widehat{\gamma(\mm{\theta})}=\frac{1}{b} \sum_{i=1}^{b}
I_{[0,\alpha]}(p(z_i)), \label{eq:approxpower}
\end{equation}
where $I$ is the indicator function having value 1 if $p(z_i)\leq\alpha$ and 0 otherwise. 
That is, the power (or test size) is estimated
as the fraction of $p$-values below $\alpha$ for the $b$ independent random draws from the
parameter vector $\theta$.

\section{A study on test size and power}\label{sec:power}
We will now investigate size and power using $p$-values from the
asymptotic approximations and
the exact conditional enumeration for the seven statistics from
Section~\ref{statistics} in various settings.

\subsection{Set-up}
In most genetic association studies the number $n_1$ of cases does not exceed the
number $n_2$ of controls. We will also make this assumption. We consider both balanced and unbalanced designs. For $n_1$ equal to 500 or 1000, we consider $n_2$ to be 1, 2, 3, 4, 5 times $n_1$. For $n_1=5000$, we consider $n_2$  to be 1, 2, 3 times $n_1$. This gives in total $13$ sample sizes $(n_1,n_2)$ to consider.

Our data generation procedure is inspired by \cite{JooRobustWellcome2009}.
We have studied a disease prevalence of 10\%, since genetic association studies
in general are designed to target common diseases. 
Most GWA studies are based on single nucleotide polymorphisms (SNPs)
with minor allele frequency (MAF) at least 5\%. We have chosen a MAF of 10\%, and we only consider
populations under Hardy--Weinberg equilibrium, which gives genotype
frequencies $g_0=(1-\text{MAF})^2=0.81$, \ $g_1=2\text{MAF}(1-\text{MAF})=0.18$, \ $g_2=\text{MAF}^2=0.01$. 
We assume that the
minor allele is also the disease allele.
For each situation under study we calculate $\theta=(p_0,p_1,p_2,q_0,q_1,q_2)$ based on the
formulas in Section~\ref{sec:genmod}, and draw data based on the probability distribution~\eqref{eq:probZ}. 
                    
Under the null hypothesis of equal penetrances for the genotypes, we
draw $4 \cdot 10^9$ samples from the study
population. For this number of simulated samples we will for a valid test at significance level
$5\cdot 10^{-8}$ get an approximate 95\% confidence
interval for the test size that have half-length $1.96\cdot \sqrt{5\cdot 10^{-8} \cdot (1-5\cdot
    10^{-8})/(4\cdot 10^9)}=7\cdot 10^{-9}$.

We also study populations under alternative hypotheses, as presented in
Section~\ref{sec:genmod}. Parameters chosen for the alternative hypotheses are the
genotype relative risk $\lambda_2=1.1$, 1.2, 1.5, 2 and the genetic
model parameter $\delta=0$, $1/4$, $1/2$, $3/4$, 1. The genotype relative risk $\lambda_1$ is
calculated from $\delta$ and $\lambda_2$ using~\eqref{eq:H1}. 

For all the
possible combinations for $\lambda_2$ and $\delta$ (as given above),
and for each of the 13
sample sizes we drew one million random samples from the
population described by these parameter values. This resulted in
$4\cdot 5\cdot 13$=260
situations. With
this number of samples we will for a true power of 
0.8 achieve an approximate 95\% confidence interval for the power with
half-length $1.96\cdot \sqrt{0.8\cdot (1-0.80)/(1\cdot 10^6)}=8
\cdot 10^{-4}$.

For each sample drawn we calculated the asymptotic and the
exact conditional enumeration
$p$-values for the seven test statistics (CATT$_{1/2}$, Pearson, MIN2,
MAX3, CMAX, CLRT and MERT) under study, and estimated the
test size or power as the relative number of
$p$-values falling below the significance levels investigated~\eqref{eq:approxpower}.

For evaluation of test size and power we present results for
significance levels $\alpha$ in the range $5 \cdot 10^{-2}$ to $5\cdot
10^{-8}$.

\subsection{Computational details}
The numerical calculations were performed in the \CC\ language of the GNU Compiler Collection 4.4.3. For generating multinomial vectors, making statistical distribution calculations and numerical integration, the GNU Scientific Library 1.13 was used. To reduce computation time, the parallel language extension OpenMP was used to distribute the generation of tables and subsequent statistical calculations among several threads operating on different processors.

For the asymptotic method, power calculations in the case of $\text{CATT}_{1/2}$, Pearson, MIN2 and MERT were done by comparing test statistics for simulated data with critical values, which is faster than calculating $p$-values explicitly. In the case of MAX3, CMAX and CLRT, $p$-values depend on estimated parameters, and had to be calculated for each simulated table.

For the exact conditional enumeration method, to avoid numerical
overflow, the hypergeometric probabilities~\eqref{eq:c} were calculated
by adding logarithms of factorials and then taking the
antilogarithm. To gain speed, the $\ln l!$ were computed once and
tabulated for $l=0$, 1, \ldots, $N$. If, for a simulated table
$(x_0,x_1,x_2,y_0,y_1,y_2)$, during the evaluation of the exact
conditional enumeration $p$-value~\eqref{eq:Cpvalue}, the sum for all
seven statistics $T$ had exceeded the highest significance level
considered, 0.05, the evaluation was aborted, since the table would
then not contribute to the power. Also, to speed up the time for a
possible early abortion of the summation \eqref{eq:Cpvalue}, the
summation was started at tables potentially having a high conditional
probability~\eqref{eq:c}, namely those having $x_0$ as the upper left
entry. Then tables having $x_0+1$ as the upper left entry were
considered and so on upwards, and thereafter the process was repeated
going from $x_0-1$ downwards. 

For the largest sample size considered in the simulation study,
$(5000,15000)$, the maximum number of summands in the calculation of the
exact conditional enumerations $p$-value is 12.5 million. This
maximum number will occur for tables having equal or nearly equal column margins. However, our choice of $\text{MAF}=0.1$ will with very low probability give such balanced column margins. 
On a $4\times6$-core Xeon 2.67 GHz computer (Intel CPU)
running Linux (Ubuntu 10.4), using one tread, the computation of exact
conditional enumeration $p$-values for 1000 tables with sample size $n_1=5000$ and $n_2=15000$ drawn from the null hypothesis took under 14 seconds. The corresponding computation for 1000 such tables drawn under an alternative
hypothesis having power near 100\% for all test statistics, took under two
minutes. 
Calculation of the exact conditional enumeration $p$-value is
faster when tables are drawn from the true null hypothesis than when
tables are drawn from the alternative
hypothesis. This is due to the fact that $p$-value calculated
for tables generated under the true null are in general larger than
$p$-values for tables generated under the alternative hypothesis, and
that we abort the calculations when the $p$-value exceeds 0.05.

For smaller sample sizes, computations are much faster. For $n_1=1000$, $n_2=1000$ the timings are 2~seconds for the null hypothesis and 0.3 seconds for the alternative.

\subsection{Effect of significance level and sample size}
In Section~\ref{sec:valid} we pointed out that there is no guarantee that asymptotic
methods preserves the test size, while the exact conditional methods
are always valid by construction. However, for $\alpha=0.05$ we found
that the asymptotic methods for nearly all test statistics (except CLRT) kept the test size for all sample sizes
investigated. With the exception of CLRT we get an increasing degree of
mismatch between the observed and the nominal level for the asymptotic methods when $\alpha$ 
decreases to $5\cdot10^{-8}$. Worst are MAX3 and CMAX for small and 
unbalanced designs, with test size up to 20 times the nominal level. For 
low significance levels, MAX3, Pearson, MIN2, MERT and CMAX keeps under 1.2 times 
the nominal level only for balanced designs and for designs having the 
number of controls twice the number of cases. $\text{CATT}_{1/2}$ fares 
a little better, but at worst has size twice the nominal level, and CLRT 
is always within 1.5 times the level. In Table~\ref{tab:H0bal} we present estimated test sizes for
balanced sample sizes for the asymptotic and exact conditional methods
for all seven test statistics for three values of the significance
level (low, $5\cdot 10^{-2}$; intermediate, $5 \cdot 10 ^{-5}$;
high, $5\cdot 10 ^{-8}$). To emphasize the need to use exact
conditional methods instead of asymptotic methods for low significance
levels and unbalanced sample sizes Table~\ref{tab:H0unbal} gives
test sizes for significance level $5\cdot 10^{-8}$ for
the asymptotic method.

\begin{sidewaystable}
\caption{Scaled test size for test statistics, methods (A asymptotic, C
 exact conditional), balanced sample sizes and selected significance
 levels $\alpha$. The test sizes shown are scaled by multiplying by $5/\alpha$, so that $5.00$ will
 give an exact test. The test size estimate is based on $4\cdot 10^9$
 simulations, giving 95\% confidence interval half-lengths of $6.8\cdot 10 ^{-6}$, \ $2.2
 \cdot 10^{-7}$, \ $6.9 \cdot 10^{-9}$ for true test sizes $5\cdot
 10^{-2}$, \ $5\cdot 10^{-5}$, \ $5 \cdot 10^{-8}$, respectively}
\label{tab:H0bal}
\begin{center}
\begin{tabular}{rrrrrrrrrrrrrrrrrrrrrrr}
 \hline
&& &\multicolumn2c{$\text{CATT}_{1/2}$} && \multicolumn2c{Pearson} && \multicolumn2c{MIN2} &&
  \multicolumn2c{MAX3} && \multicolumn2c{CMAX} && \multicolumn2c{CLRT} && \multicolumn2c{MERT}\\ 
 \cline{4-5}\cline{7-8}\cline{10-11}\cline{13-14}\cline{16-17}\cline{19-20}\cline{22-23}
 \multicolumn1c{$n_1$} & \multicolumn1c{$n_2$} & 
\multicolumn1c{$\alpha$} &
\multicolumn1c{A} & \multicolumn1c{C} && \multicolumn1c{A} &
  \multicolumn1c{C} && \multicolumn1c{A} & \multicolumn1c{C} && \multicolumn1c{A} & \multicolumn1c{C} && \multicolumn1c{A} &
  \multicolumn1c{C} && \multicolumn1c{A} & \multicolumn1c{C} && \multicolumn1c{A} & \multicolumn1c{C}\\
 \hline
 500 & 500 & $5 \cdot 10 ^{-2}$  & 5.00 & 4.21 &   & 4.78 & 4.91 &   & 4.77 & 4.66 &   & 4.75 & 4.14 &   & 4.62 & 4.73 &   & 5.43 & 4.55 &   & 4.93 & 4.81 \\ 
   &  &  $5\cdot 10^{-5}$& 4.53 & 3.66 &   & 2.28 & 4.82 &   & 3.24 & 4.13 &   & 2.61 & 4.20 &   & 2.04 & 4.64 &   & 5.38 & 4.59 &   & 2.08 & 4.46 \\ 
 && $5\cdot 10^{-8}$ & 3.15 & 2.95 &   & 1.20 & 4.80 &   & 2.02 & 3.42 &   & 1.85 & 3.42 &   & 1.05 & 3.98 &   & 3.02 & 3.88 &   & 0.15 & 3.98 \\ 
  1000 & 1000 & $5 \cdot 10 ^{-2}$ & 5.00 & 4.42 &   & 4.89 & 4.96 &   & 4.90 & 4.76 &   & 4.85 & 4.39 &   & 4.64 & 4.80 &   & 5.00 & 4.80 &   & 4.98 & 4.91 \\ 
  &&$5\cdot 10^{-5}$ & 4.76 & 4.02 &   & 3.06 & 4.96 &   & 3.73 & 4.48 &   & 3.10 & 4.45 &   & 2.75 & 4.87 &   & 6.19 & 4.70 &   & 3.36 & 4.80 \\ 
  &&$5 \cdot 10^{-8}$ & 4.83 & 4.35 &   & 1.92 & 5.22 &   & 3.20 & 4.67 &   & 2.77 & 4.62 &   & 1.77 & 5.10 &   & 6.42 & 4.45 &   & 1.07 & 4.85 \\ 
  5000 & 5000 & $5 \cdot 10 ^{-2}$ & 5.00 & 4.73 &   & 4.98 & 4.99 &   & 4.98 & 4.90 &   & 4.97 & 4.71 &   & 4.76 & 4.93 &   & 4.81 & 4.93 &   & 5.00 & 4.98 \\ 
  && $5 \cdot 10^{-5}$ &4.95 & 4.53 &   & 4.52 & 4.99 &   & 4.68 & 4.82 &   & 4.57 & 4.58 &   & 4.29 & 4.95 &   & 4.95 & 4.92 &   & 4.65 & 4.96 \\ 
  && $5\cdot 10^{-8}$ & 4.53 & 3.80 &   & 4.10 & 5.03 &   & 3.70 & 4.05 &   & 3.58 & 4.33 &   & 3.20 & 4.67 &   & 4.95 & 4.70 &   & 4.12 & 5.00 \\ 
   \hline
\end{tabular}
\end{center}
\end{sidewaystable}

\newcommand{\hh}{\hphantom0}

\begin{table}
\caption{Scaled test size for test statistics, balanced and
  unbalanced sample sizes for the asymptotic method  for significance level $\alpha=5 \cdot 10
  ^{-7}$. The test sizes shown are scaled by multiplying by 5/$\alpha$, so that $5.00$ will
  give an exact test. $4\cdot 10^9$ simulations are
  run, giving 95\% confidence interval half-lengths of $2.2 \cdot 10^{-8}$ for true test sizes $5\cdot
  10^{-7}$.}
\label{tab:H0unbal}
\begin{center}
\begin{tabular}{rrccccccc}
 \hline
 \multicolumn1c{$n_1$} & \multicolumn1c{$n_2$} &$\text{CATT}_{1/2}$ & Pearson & MIN2 &
  MAX3 &CMAX &CLRT & MERT\\ 
\hline
500 &500  & 3.81 & \hh1.61 & \hh2.55 & \hh2.28 & \hh1.50 & 3.66 & \hh0.54 \\ 
  &1000  & 5.14 & \hh4.02 & \hh4.37 & \hh4.05 & \hh4.20 & 5.05 & \hh4.57 \\ 
  & 1500  & 6.01 & 14.81 & 11.78 & 17.52 & 15.12 & 4.32 & 11.80 \\ 
  & 2000  & 6.89 & 28.76 & 21.76 & 33.46 & 32.90 & 3.99 & 18.44 \\ 
  & 2500  & 7.60 & 42.75 & 31.05 & 50.29 & 46.66 & 4.01 & 23.97 \\ 
  1000& 1000  & 4.55 & \hh2.19 & \hh3.30 & \hh2.71 & \hh1.95 & 6.83 & \hh1.78 \\ 
  & 2000  & 5.15 & \hh5.76 & \hh5.46 & \hh6.12 & \hh5.75 & 5.97 & \hh5.46 \\ 
  & 3000  & 5.35 & 11.84 & \hh9.54 & 12.33 & 12.82 & 5.31 & \hh9.04 \\ 
  & 4000  & 5.89 & 17.95 & 13.99 & 21.49 & 20.25 & 5.14 & 12.17 \\ 
  & 5000  & 6.22 & 23.74 & 17.95 & 26.95 & 26.23 & 5.13 & 14.90 \\ 
  5000& 5000  & 4.85 & \hh4.01 & \hh4.29 & \hh4.06 & \hh3.75 & 5.05 & \hh4.13 \\ 
  & 10000  & 5.10 & \hh5.21 & \hh5.18 & \hh5.45 & \hh5.42 & 4.97 & \hh5.31 \\ 
  &15000  & 5.23 & \hh6.84 & \hh6.30 & \hh7.27 & \hh7.23 & 5.11 & \hh6.21 \\ 
\hline
 \end{tabular}
\end{center}
\end{table}

Turning to the power study, we find that the power increases with increasing sample sizes
and with increasing significance levels. This can be seen in
Table~\ref{tab:effectsign} (the effect of significance level) and
\ref{tab:effectn1n2} (the effect of sample size).
Keeping the total sample size fixed the highest power is observed for
balanced sample sizes and decreases with increasing degree of
unbalance. This can be observed (Table~\ref{tab:effectn1n2}) for the sample size combinations $(n_1,n_2)=(500, 1500)$ and
$(1000,1000)$, both giving total sample size 2000, and this is also the case for $(500,2500)$ and
$(1000,2000)$, with total sample size 3000.

\begin{sidewaystable}
\caption{Power: Effect of test statistic, method (A asymptotic, C
  exact conditional) and significance level ($\alpha$). Sample size is
  $(n_1,n_2)=(5000,5000)$, genetic model is additive ($\delta=0.5$)
  and genetic relative risk $\lambda_2=1.5$.}
\label{tab:effectsign}
\begin{center}
\begin{tabular}{crrrrrrrrrrrrrrrrrrrr}
 \hline
 & \multicolumn2c{$\text{CATT}_{1/2}$} && \multicolumn2c{Pearson} && \multicolumn2c{MIN2} &&
  \multicolumn2c{MAX3} && \multicolumn2c{CMAX} && \multicolumn2c{CLRT} && \multicolumn2c{MERT}\\ 
 \cline{2-3}\cline{5-6}\cline{8-9}\cline{11-12}\cline{14-15}\cline{17-18}\cline{20-21}
 $\alpha$ & \multicolumn1c{A} & \multicolumn1c{C} && \multicolumn1c{A} &
  \multicolumn1c{C} && \multicolumn1c{A} & \multicolumn1c{C} && \multicolumn1c{A} & \multicolumn1c{C} && \multicolumn1c{A} &
  \multicolumn1c{C} && \multicolumn1c{A} & \multicolumn1c{C} && \multicolumn1c{A} & \multicolumn1c{C}\\
 \hline
 $5\cdot10^{-2}$ & 100.0 & 100.0 && 99.9 & 99.9 && 100.0 & 99.9 && 99.9 & 99.9 && 99.9 & 99.9 && 99.9 & 99.9 && 99.8 & 99.8 \\ 
 $5\cdot10^{-3}$ & 99.5 & 99.5 && 98.8 & 98.8 && 99.3 & 99.3 && 99.2 & 99.2 && 99.1 & 99.2 && 99.1 & 99.1 && 97.8 & 97.8 \\ 
 $5\cdot10^{-4}$ & 97.2 & 97.1 && 94.6 & 94.7 && 96.4 & 96.4 && 96.0 & 96.1 && 95.7 & 95.9 && 95.7 & 95.7 && 90.8 & 90.9 \\ 
 $5\cdot10^{-5}$ & 91.0 & 90.6 && 85.3 & 85.7 && 89.1 & 89.1 && 88.4 & 88.7 && 87.4 & 88.0 && 87.5 & 87.6 && 77.4 & 77.8 \\ 
 $5\cdot10^{-6}$ & 79.6 & 79.1 && 71.0 & 71.8 && 76.7 & 76.8 && 75.8 & 76.4 && 74.1 & 75.3 && 74.3 & 74.3 && 59.4 & 60.2 \\ 
 $5\cdot10^{-7}$ & 64.4 & 63.7 && 54.0 & 55.4 && 60.8 & 61.2 && 59.8 & 60.9 && 57.7 & 59.5 && 58.0 & 58.0 && 41.0 & 42.1 \\ 
 $5\cdot10^{-8}$ & 47.6 & 47.0 && 37.5 & 39.2 && 44.0 & 44.6 && 43.1 & 44.5 && 41.0 & 43.1 && 41.3 & 41.2 && 25.5 & 26.7 \\ 
 \hline
\end{tabular}
\end{center}
\end{sidewaystable}

\begin{sidewaystable}
\caption{Power: Effect of test statistic, method (A asymptotic, C exact conditional) and sample size ($n_1$ cases, $n_2$ controls). The genetic model is
semi-dominant ($\delta=3/4$), 
genetic relative risk $\lambda_2=1.5$ and 
significance level $\alpha=5\cdot10^{-5}$.}
\label{tab:effectn1n2}
\begin{center}
\begin{tabular}{rrrrrrrrrrrrrrrrrrrrrr}
 \hline
 && \multicolumn2c{$\text{CATT}_{1/2}$} && \multicolumn2c{Pearson} && \multicolumn2c{MIN2} &&
  \multicolumn2c{MAX3} && \multicolumn2c{CMAX} && \multicolumn2c{CLRT} && \multicolumn2c{MERT}\\ 
 \cline{3-4}\cline{6-7}\cline{9-10}\cline{12-13}\cline{15-16}\cline{18-19}\cline{21-22}
 \multicolumn1c{$n_1$} & \multicolumn1c{$n_2$} & \multicolumn1c{A} & \multicolumn1c{C} && \multicolumn1c{A} &
  \multicolumn1c{C} && \multicolumn1c{A} & \multicolumn1c{C} && \multicolumn1c{A} & \multicolumn1c{C} && \multicolumn1c{A} &
  \multicolumn1c{C} && \multicolumn1c{A} & \multicolumn1c{C} && \multicolumn1c{A} & \multicolumn1c{C}\\
 \hline
 500 &500 & 3.9 & 3.5 &  & 2.3 & 3.3 &  & 3.2 & 3.5 &  & 3.1 & 3.9 &  & 2.7 & 3.9 &  & 3.0 & 2.8 &  & 1.4 & 2.0 \\ 
  & 1000 & 9.9 & 9.2 &  & 6.9 & 7.1 &  & 8.6 & 8.4 &  & 8.4 & 8.4 &  & 7.7 & 7.9 &  & 6.7 & 6.5 &  & 4.4 & 4.6  \\ 
  & 1500  & 14.1 & 12.9 &  & 10.4 & 8.6 &  & 12.5 & 10.6 &  & 12.3 & 10.1 &  & 11.4 & 9.3 &  & 9.2 & 9.5 &  & 6.6 & 5.8  \\ 
   & 2000& 16.9 & 15.6 &  & 13.0 & 9.5 &  & 15.3 & 11.9 &  & 15.1 & 10.9 &  & 14.1 & 10.1 &  & 11.1 & 11.5 &  & 8.3 & 6.7   \\ 
  &  2500 & 19.0 & 17.3 &  & 14.7 & 10.0 &  & 17.2 & 12.7 &  & 17.0 & 11.3 &  & 16.0 & 10.5 &  & 12.3 & 12.8 &  & 9.4 & 7.2   \\ 
  1000 &1000 & 21.6 & 20.5 &  & 15.6 & 17.9 &  & 19.1 & 19.9 &  & 19.1 & 21.1 &  & 17.2 & 20.1 &  & 17.8 & 16.8 &  & 8.7 & 9.9   \\ 
  & 2000  & 42.4 & 41.2 &  & 34.6 & 34.6 &  & 39.3 & 38.7 &  & 39.3 & 38.9 &  & 36.9 & 37.3 &  & 34.8 & 33.7 &  & 20.9 & 21.2   \\ 
  & 3000 & 52.8 & 51.3 &  & 44.9 & 42.1 &  & 49.7 & 47.2 &  & 49.8 & 46.4 &  & 47.4 & 44.4 &  & 44.0 & 42.5 &  & 28.0 & 26.9   \\ 
  &4000  & 58.7 & 57.3 &  & 51.2 & 46.5 &  & 55.9 & 52.1 &  & 56.0 & 50.7 &  & 53.7 & 48.7 &  & 49.5 & 47.8 &  & 32.7 & 30.5   \\ 
  & 5000 & 62.5 & 61.0 &  & 55.2 & 49.4 &  & 59.8 & 55.0 &  & 59.9 & 53.4 &  & 57.6 & 51.3 &  & 53.1 & 51.2 &  & 35.8 & 32.8   \\ 
  5000 & 5000  & 99.9 & 99.9 &  & 99.9 & 99.9 &  & 99.9 & 99.9 &  & 99.9 & 99.9 &  & 99.9 & 99.9 &  & 99.9 & 99.9 &  & 98.0 & 98.0   \\ 
  & 10000  & 100.0 & 100.0 &  & 100.0 & 100.0 &  & 100.0 & 100.0 &  & 100.0 & 100.0 &  & 100.0 & 100.0 &  & 100.0 & 100.0 &  & 99.9 & 99.9   \\ 
  & 15000 & 100.0 & 100.0 &  & 100.0 & 100.0 &  & 100.0 & 100.0 &  & 100.0 & 100.0 &  & 100.0 & 100.0 &  & 100.0 & 100.0 &  & 100.0 & 100.0   \\ 
 \hline
\end{tabular}
\end{center}
\end{sidewaystable}

\subsection{Effect of genetic model}
Since the asymptotic methods, with CLRT as the exception, are in general not valid, we base our
discussion on comparing power based on the exact conditional enumeration $p$-values. But, the power of the invalid asymptotic methods will in
general not be substantially larger than the power of the valid exact conditional
methods, which is seen in Tables~\ref{tab:effectsign}--\ref{tab:tab80}.

We have chosen to only study monotone genetic models, and we find that
the effect of
genetic model seem to be similar for all sample sizes and significance
levels. 
Results for $(n_1,n_2)=(5000,15000)$ and $\alpha=5\cdot10^{-6}$ are shown in Table~\ref{tab:effectgenmodel}.

For the recessive model ($\delta=0$) the CMAX and MAX3 methods
performs the best (with very similar powers). 
The $\text{CATT}_{1/2}$ performs poorly for the recessive model, as compared to all the other test statistics studied. The most extreme situation observed was for sample size
(5000,15000) for $\lambda_2=2$ and $\alpha=5\cdot 10^{-6}$, where
$\text{CATT}_{1/2}$ gives a power of 4.6\% while the CMAX gives a power
of 88.4\% (Table \ref{tab:effectgenmodel}, fourth row from the top).

For the semi-recessive model ($\delta=1/4$) MERT gives the best
performance (Table \ref{tab:effectgenmodel},
rows 5--8). 

For the additive model ($\delta=1/2$) the $\text{CATT}_{1/2}$ performs
the best. This is also true for the semi-dominant ($\delta=3/4$)
and the dominant ($\delta=1$) models. The MERT test statistic has for these
three genetic models lower power than the other test statistics. The other
test statistics (Pearson, MIN2, MAX3, CMAX and CLRT) have comparable powers,
in most cases slightly lower than CATT$_{1/2}$ and higher than MERT (Table~\ref{tab:effectgenmodel}, lower part; Table~\ref{tab:effectsign} for the additive model).

\begin{sidewaystable}
\caption{Power: Effect of test statistic, method (A asymptotic, C
  exact conditional), genetic model ($\delta=0$ recessive, $\delta=1$
  dominant) and genetic relative risk ($\lambda_2$). Sample size is
  $(n_1,n_2)=(5000,15000)$ and significance level
  $\alpha=5\cdot10^{-6}$.}
\label{tab:effectgenmodel}
\begin{center}
\begin{tabular}{llrrrrrrrrrrrrrrrrrrrr}
 \hline
 && \multicolumn2c{$\text{CATT}_{1/2}$} && \multicolumn2c{Pearson} && \multicolumn2c{MIN2} &&
  \multicolumn2c{MAX3} && \multicolumn2c{CMAX} && \multicolumn2c{CLRT} && \multicolumn2c{MERT}\\ 
 \cline{3-4}\cline{6-7}\cline{9-10}\cline{12-13}\cline{15-16}\cline{18-19}\cline{21-22}
 \multicolumn1c{$\delta$} & \multicolumn1c{$\lambda_2$} & \multicolumn1c{A} & \multicolumn1c{C} && \multicolumn1c{A} &
  \multicolumn1c{C} && \multicolumn1c{A} & \multicolumn1c{C} && \multicolumn1c{A} & \multicolumn1c{C} && \multicolumn1c{A} &
  \multicolumn1c{C} && \multicolumn1c{A} & \multicolumn1c{C} && \multicolumn1c{A} & \multicolumn1c{C}\\
 \hline
 0 & 1.1 & 0.0 & 0.0 && 0.0 & 0.0 && 0.0 & 0.0 && 0.0 & 0.0 && 0.0 & 0.0 && 0.0 & 0.0 && 0.0 & 0.0 \\ 
 & 1.2 & 0.0 & 0.0 && 0.1 & 0.1 && 0.0 & 0.0 && 0.1 & 0.1 && 0.1 & 0.1 && 0.0 & 0.0 && 0.0 & 0.0 \\ 
 & 1.5 & 0.1 & 0.1 && 7.6 & 7.2 && 6.3 & 6.1 && 8.6 & 7.5 && 8.2 & 7.7 && 5.4 & 5.4 && 2.8 & 2.7 \\ 
 & 2 & 4.7 & 4.6 && 88.2 & 87.6 && 86.2 & 85.7 && 89.8 & 88.5 && 89.1 & 88.4 && 84.6 & 84.6 && 60.5 & 59.9 \\ 
 0.25 & 1.1 & 0.0 & 0.0 && 0.0 & 0.0 && 0.0 & 0.0 && 0.0 & 0.0 && 0.0 & 0.0 && 0.0 & 0.0 && 0.0 & 0.0 \\ 
 & 1.2 & 0.3 & 0.2 && 0.2 & 0.2 && 0.3 & 0.2 && 0.2 & 0.2 && 0.3 & 0.3 && 0.2 & 0.2 && 0.4 & 0.4 \\ 
 & 1.5 & 34.9 & 34.3 && 33.7 & 32.5 && 35.3 & 34.3 && 30.8 & 29.1 && 37.2 & 36.0 && 33.4 & 33.5 && 41.5 & 40.8 \\ 
 & 2 & 99.9 & 99.9 && 99.9 & 99.9 && 100.0 & 100.0 && 99.9 & 99.9 && 100.0 & 100.0 && 99.9 & 99.9 && 100.0 & 100.0 \\ 
 0.5 & 1.1 & 0.1 & 0.1 && 0.1 & 0.1 && 0.1 & 0.1 && 0.1 & 0.1 && 0.1 & 0.1 && 0.1 & 0.1 && 0.1 & 0.1 \\ 
 & 1.2 & 4.3 & 4.2 && 2.7 & 2.5 && 3.6 & 3.4 && 3.3 & 3.0 && 3.2 & 3.0 && 2.8 & 2.8 && 2.5 & 2.4 \\ 
 & 1.5 & 98.3 & 98.2 && 96.8 & 96.5 && 97.8 & 97.7 && 97.6 & 97.4 && 97.4 & 97.2 && 97.0 & 97.0 && 92.1 & 91.8 \\ 
 & 2 & 100.0 & 100.0 && 100.0 & 100.0 && 100.0 & 100.0 && 100.0 & 100.0 && 100.0 & 100.0 && 100.0 & 100.0 && 100.0 & 100.0 \\ 
 0.75 & 1.1 & 0.6 & 0.5 && 0.3 & 0.3 && 0.5 & 0.4 && 0.4 & 0.4 && 0.4 & 0.4 && 0.3 & 0.3 && 0.2 & 0.2 \\ 
 & 1.2 & 25.6 & 25.1 && 19.7 & 18.7 && 23.2 & 22.3 && 23.4 & 22.1 && 21.5 & 20.5 && 20.0 & 20.0 && 10.2 & 9.9 \\ 
 & 1.5 & 100.0 & 100.0 && 100.0 & 100.0 && 100.0 & 100.0 && 100.0 & 100.0 && 100.0 & 100.0 && 100.0 & 100.0 && 99.9 & 99.8 \\ 
 & 2 & 100.0 & 100.0 && 100.0 & 100.0 && 100.0 & 100.0 && 100.0 & 100.0 && 100.0 & 100.0 && 100.0 & 100.0 && 100.0 & 100.0 \\ 
 1 & 1.1 & 2.3 & 2.2 && 1.6 & 1.5 && 2.0 & 1.9 && 2.1 & 1.8 && 1.7 & 1.6 && 1.5 & 1.5 && 0.7 & 0.6 \\ 
 & 1.2 & 64.7 & 64.1 && 59.9 & 58.5 && 63.2 & 62.0 && 64.6 & 62.9 && 61.6 & 60.2 && 59.7 & 59.8 && 27.7 & 27.1 \\ 
 & 1.5 & 100.0 & 100.0 && 100.0 & 100.0 && 100.0 & 100.0 && 100.0 & 100.0 && 100.0 & 100.0 && 100.0 & 100.0 && 100.0 & 100.0 \\ 
 & 2 & 100.0 & 100.0 && 100.0 & 100.0 && 100.0 & 100.0 && 100.0 & 100.0 && 100.0 & 100.0 && 100.0 & 100.0 && 100.0 & 100.0 \\ 
 \hline
\end{tabular}
\end{center}
\end{sidewaystable}

\subsection{General findings}

From the results of our power study we may divide the test statistics into four groups based on their
overall performance. (i) The $\text{CATT}_{1/2}$
has very good performance for all models other than the recessive, (ii)
MERT performs well for the semi-recessive model, but else has a less good
performance, (iii) the three test statistics MAX3, CMAX and CLRT have very similar
performance and give good results for all genetic models, and 
lastly, (iv) Pearson and MIN2 also have very similar performance, in
general slightly less powerful than the previous group, but is
also known to work well for non-monotone genetic models \citep{JooRobustWellcome2009}.


In Table~\ref{tab:tab80} we present powers for $\alpha=5\cdot 10^{-8}$
for all the sample sizes under study. For each sample size we have chosen the genetic model and effect
size that give power (over all test statistics) closest to 80\%. 
These results are influenced by a selection bias due to the fact that for small
sample sizes only the dominant models with large effects sizes will
achieve power near 80\%, and for large sample sizes power near 80\%
will be achieved for additive to recessive models. Taking this into mind, we see that the
general results presented above are reflected in this table. To
summarize, the exact conditional enumeration method and the MAX3 test
statistics is the most powerful for small balanced and slightly
unbalanced sample sizes, $(500,500)$, $(500,1000)$ and $(1000,1000)$, for the dominant model. 
For $(500,1000)$, the MAX3 test statistics (exact conditional
enumeration) gives 46.6 percentage points higher power than the MERT test
statistic.
For the unbalanced sample sizes, $(500,1500)$, $(500,2000)$, $(500,2500)$, $(1000,
2000)$, $(1000,3000)$, $(1000,4000)$, $(1000,5000)$, the asymptotic method for
the MAX3 test statistic is the most powerful for the dominant and
semi-dominant model. This is not surprising
since the asymptotic MAX3 shows large violations for these unbalanced
sample sizes. For this reason, we do not recommend using the asymptotic
method for these test statistics for unbalanced sample sizes. Only
considering the exact conditional enumeration method for these
unbalanced sample sizes, the best performance is found for the
$\text{CATT}_{1/2}$ test statistics. For the largest sample sizes, MERT
performs the best for the semi-recessive model for $(5000,5000)$,
while the $\text{CATT}_{1/2}$ performs the best for the large unbalance
sample sizes $(5000,10000)$, $(5000,15000)$ under the additive model.
 
\begin{sidewaystable}
\caption{Power: Effect of test statistic, method (asymptotic, A, and
  exact conditional, C) and sample size ($n_1$ cases, $n_2$ controls) for
  significance level $\alpha=5\cdot10^{-8}$. Values of the genetic
  model $\delta$ and
  $\lambda_2$ are chosen to give power closest to 80\%. For each
  sample size the most powerful combination of method and test
  statistic is given in bold face, while the most powerful exact
  conditional enumeration method and test statistic is given in italic.}
\label{tab:tab80}
\begin{center}
\begin{tabular}{rrlrrrrrrrrrrrrrrrrrrrrr}
 \hline
&& && \multicolumn2c{$\text{CATT}_{1/2}$} && \multicolumn2c{Pearson} && \multicolumn2c{MIN2} &&
  \multicolumn2c{MAX3} && \multicolumn2c{CMAX} && \multicolumn2c{CLRT} && \multicolumn2c{MERT}\\ 
 \cline{5-6}\cline{8-9}\cline{11-12}\cline{14-15}\cline{17-18}\cline{20-21}\cline{23-24}
 \multicolumn1c{$n_1$} & \multicolumn1c{$n_2$} & 
\multicolumn1c{$\delta$} &
\multicolumn1c{$\lambda_2$} &
\multicolumn1c{A} & \multicolumn1c{C} && \multicolumn1c{A} &
  \multicolumn1c{C} && \multicolumn1c{A} & \multicolumn1c{C} && \multicolumn1c{A} & \multicolumn1c{C} && \multicolumn1c{A} &
  \multicolumn1c{C} && \multicolumn1c{A} & \multicolumn1c{C} && \multicolumn1c{A} & \multicolumn1c{C}\\
 \hline
  500 & 500 & 1 & 2.0 & 36.8 & 36.0 &  & 32.3 & 39.2 &  & 35.3 & 38.0
  &  & 37.5 & 41.2 &  & 34.2 & \textbf{\emph{41.4}} &  & 36.0 & 37.3 &  & 7.5 & 14.0 \\ 
   & 1000 & 1 & 2.0 & 74.1 & 72.3 &  & 71.6 & 72.6 &  & 73.8 & 73.6 &
   & 75.7 & \textbf{\emph{ 76.1}} &  & 73.1 & 73.8 &  & 69.8 & 69.7 &  & 29.5 & 29.4 \\ 
 & 1500 & 1 & 2.0 & 86.2 & \emph{ 83.8} &  & 84.9 & 78.2 &  & 86.2 &
 80.9 &  & \textbf{87.6} & 82.1 &  & 85.8 & 79.0 &  & 81.9 & 82.7 &  & 43.8 & 36.0 \\ 
   & 2000 & 1 & 2.0 & 91.0 & \emph{ 88.7} &  & 90.2 & 79.9 &  & 91.1 &
   83.0 &  & \textbf{92.1} & 82.8 &  & 90.9 & 80.1 &  & 87.3 & 88.0 &  & 52.4 & 39.8 \\ 
  &2500 & 1 & 2.0 & 93.4 & \emph{ 91.2} &  & 92.9 & 81.0 &  & 93.6 &
  84.0 &  & \textbf{94.3} & 82.8 &  & 93.4 & 81.1 &  & 90.1 & 90.8 &  & 57.9 & 42.4 \\ 
 1000 & 1000 & 0.75 & 2.0 & 68.4 & 67.6 &  & 62.0 & 67.1 &  & 66.3 &
 68.1 &  & 67.7 & \textbf{\emph{ 70.8} }&  & 64.7 & 70.3 &  & 65.8 & 63.6 &  & 28.7 & 34.8 \\ 
 & 2000 & 0.75 & 2.0 & \textbf{93.8} & \emph{ 93.4} &  & 91.8 & 91.4 &  &
 93.3 & 92.8 &  & \textbf{93.8} & 93.2 &  & 92.8 & 92.3 &  & 91.6 & 91.2 &  & 65.1 & 64.1 \\ 
 & 3000 & 0.75 & 2.0 & 97.8 & \emph{ 97.4} &  & 97.0 & 95.2 &  & 97.6 &
 96.4 &  & \textbf{97.9} & 96.3 &  & 97.4 & 95.6 &  & 96.6 & 96.5 &  & 78.8 & 74.1 \\ 
& 4000 & 0.75 & 2.0 & \textbf{98.9} & 98.6 &  & 98.5 & 96.6 &  & 98.8 &
97.5 &  & \textbf{98.9} & 97.3 &  & 98.7 & \emph{96.8} &  & 98.1 & 98.1 &  & 85.0 & 79.0 \\ 
 & 5000 & 0.75 & 2.0 & 99.3 & \emph{ 99.1} &  & 99.0 & 97.3 &  & 99.3 &
 98.1 &  & \textbf{99.4} & 97.9 &  & 99.2 & 97.5 &  & 98.7 & 98.8 &  & 88.3 & 81.9 \\ 
  5000& 5000 & 0.25 & 2.0 & 81.5 & 81.1 &  & 78.6 & 79.7 &  & 80.7 & 81.1 &
  & 77.1 & 78.0 &  & 81.3 & 82.6 &  & 82.0 & 82.0 &  & 83.0 & {\bf
    \emph{ 83.8}} \\ 
 & 10000 & 0.50 & 1.5 & \textbf{80.3} & \emph{ 79.9} &  & 72.7 & 71.9 &  & 77.8 & 77.1 &  & 77.1 & 76.1 &  & 75.5 & 74.7 &  & 73.8 & 73.8 &  & 57.8 & 57.3 \\ 
 & 15000 & 0.50 & 1.5 & \textbf{89.6} & \emph{ 89.2} &  & 84.5 & 82.5 &  & 88.0 & 86.7 &  & 87.5 & 85.4 &  & 86.5 & 84.4 &  & 84.7 & 84.7 &  & 71.6 & 69.9 \\ 
   \hline
\end{tabular}
\end{center}
\end{sidewaystable}

\section{Discussion}\label{sec:discuss}
\paragraph{Parameter choices in the simulation study}
In the simulation study (Section \ref{sec:power}) all data have been
generated assuming that the disease prevalence is $k=0.1$, the minor
allele (disease allele) frequency is $\text{MAF}=0.1$, and that the total population is in Hardy--Weinberg equilibrium (HWE). Further, we have only studied monotone
genetic models with low to moderate effects size $\lambda_2$ and sample sizes in the order 500--5000 cases and
500--15000 controls. The conclusions to be drawn from the simulations
study are thus only valid for these situations. However, some
observations on the effect of changes to the set-up may be drawn.

The data are simulated in a case--control design.
Keeping all other parameters fixed, the effect of doubling the disease
prevalence is to double the probability of disease for each genotype. This will leave the genotype
probabilities for the cases unchanged, and will only change the
genotype probabilities for the controls slightly. We believe that the
effect of changing the disease prevalence in our study will be minor.
 
There is a straightforward effect of changing the MAF. The simulation
study used $\text{MAF}=0.1$, and lowering the MAF will, most importantly, lead to a lower
probability for the disease type homozygotes, $g_2=\text{MAF}^2$.
This will in turn lead to a greater imbalance in the expected cell
counts for the six contingency table cells, influencing the validity
of the asymptotic approximations in a negative manner.

Since we are working with test statistics that are based on genotype data, there is no need to assume
HWE. We may generate data deviating from HWE by introducing an inbreeding
coefficient when calculating genotype frequencies. In a data model
with positive inbreeding coefficient the genotype frequencies for the two groups of homozygotes will increase and the
genotype frequency for the heterozygote will decrease. 
We believe this will influence the disease homozygote group the most,
and that this in turn will lead to a better balance in the expected cell counts for the
six contingency table cells, thus, influencing the asymptotic
approximations in a positive manner.

\paragraph{Validity and asymptotic methods}
For candidate studies and intermediate follow-up studies producing a $p$-value for
each genetic marker is in general of interest, while for GWA studies
the main objective may be to provide a ranking of the genetic markers (with respect to
increasing strength of the disease--genotype association).

We first discuss ranking of test statistics and ranking of $p$-value for GWA
studies. Assume that in a case--control study with $n_1$ cases and $n_2$
controls we have studied $m$ genetic markers. These markers may in
general come from a population with different genotype frequencies for
each marker. The collected data would typically have
different column margins.  

For the test statistics $\text{CATT}_{1/2}$, Pearson, MIN2 and MERT, the asymptotic
null distribution of the test statistics does not depend upon any unknown
parameters, and also not on $n_1$ and $n_2$. This means that the rank
of the $m$ genetic markers
according to each of these test statistics will be the same as the
rank according to the corresponding asymptotic $p$-value. 
For the test
statistics MAX3, CMAX and CLRT the asymptotic null distribution is
dependent on the estimated value of the genotype frequencies
$g_0$, $g_1$, $g_2$. The ranking of the genetic markers by test statistic may differ from the
ranking by the asymptotic $p$-values, but we believe that the
difference in ranking is minor.
For the ranking of the test statistics as compared to the ranking of
the exact conditional enumeration $p$-value, the ranking of the exact
conditional enumeration $p$-value will only be the same as the ranking
of the test statistic for genetic markers with identical column
margins. In a GWA study the ranking of the genetic markers based on a test statistic will be
different from the ranking based on exact conditional enumeration
$p$-values. To which extent the rankings differ may be a topic of
further study.


When a $p$-value is calculated and used to guide the acceptance or
rejection of one or many null hypotheses we would like to advocate
using methods that produce valid $p$-values. Otherwise,
violations of the single or multiple type I error will lead to
loss of type I error control, and to optimistic power calculations.

The lack of validity of the $p$-values from asymptotic methods for low
significance levels is in general not surprising, and has for other test statistics been observed by \cite{Morris2011}.

In our simulation
study the expected cell counts under the null hypothesis were $g_i n_1$ and $g_i n_2$ for genotypes
$i=0,1,2$ for cases and controls, respectively, which gave the
smallest expected cell count of $g_2 n_1=0.01 \cdot 500=5$ over all our
simulations. Thus, for all sample sizes studied there are no expected
cell counts below 5 under the null hypothesis.

The asymptotic $p$-value calculated from the Pearson test statistic
has been studied extensively. \cite{Cochran1954} formulated the
following rule of thumb for using the Pearson test statistic
with the asymptotic chi-squared approximation for contingency tables with
more than one degree of freedom (larger than $2 \times 2$). ``If relatively few expectations
are less than 5 (say in 1 cell out of 5 or more, or 2 cells out of
10 or more), a minimum expectation of 1 is allowable in computing $\chi^2$.''
However, we found that the asymptotic $p$-value for the Pearson test
did not keep its test size, especially for unbalanced sample sizes and low
significance levels, even if all cells had expected
count at least 5. 

\cite{Wise1963} found that errors in
the chi-squared approximation in a multinomial situation are particularly small when the expected
cell counts are equal or nearly so, and that these expected cell
counts need not be large.
Our results point towards a greater effect of equality in expected cells
counts than of the numerical values of the expected cell
counts. This can be seen by comparing the test sizes in
Table~\ref{tab:H0unbal}. For $n_1=500$, observe that for the Pearson test statistics
(and also for the MIN2, MAX3, CMAX and MERT) the violations in test
size increase as $n_2$ increases from 500 to 2500. When $n_2$ increases,
the expected cell counts for the controls will increase, but the difference
between the expected cell counts between the cases and controls will
also increase. The same pattern is
seen for $n_1=1000$ as $n_2$ increases.

In addition to the Pearson test statistics, the $\text{CATT}_{1/2}$ and the MERT test
statistics follow the chi-squared and standard normal asymptotic null
distributions. 
We believe that the observations on
equality of expected cell counts for the Pearson test statistic will
also apply to the other test statistics. 

\paragraph{Environmental covariates and logistic regression}
All the methods considered in this presentation use only information
on the disease phenotype and the genotype in order to calculate test
statistics and
$p$-values. However, data on environmental covariates
may also have been collected. It is believed that complex
diseases may be the result of an interplay between genetic markers and 
environmental covariates. 

The asymptotic CATT$_s$ test can be performed by first fitting a logistic regression to
the disease status as response and the genotype as covariates, where
the value 0 is used as coding for the wild type homozygotes, 1 for the disease
homozygotes, and $s$ for the heterozygotes, and then performing an
asymptotic score test.  This strategy is easily extended to include
environmental covariates and interactions between environmental
covariates and genotype. In a simulation study \cite{Runde2013} found
that the power gain in including an environmental covariate (when
present) in a logistic regression score test 
is minimal compared to using the asymptotic CATT$_s$ 
unless the effect of the (standardized) environmental
covariate corresponds to a least an odds ratio of 5. 

Robust methods for disease--genotype association with covariates are
available. \cite{SoSham2011}
has developed a MAX3 type method combining three asymptotic logistic
regression score test. However, the validity of the method, in
particular when using low significance levels, has to our knowledge not
been investigated. Exact conditional logistic regression are available
\citep{MehtaPatel1995}, but to our knowledge exact methods have not been investigated for
robust methods with covariates.

\section{Conclusions}\label{sec:conclude}
We have studied seven test statistics that can be used to detect
association between a dichotomous phenotype (disease) and genotype.
We advocate that if you work with GWA studies,
and would like to detect all monotone genetic models (including the
recessive) with high power you should not
use the CATT$_{1/2}$ test statistics, but instead work with MAX3, CMAX
or CLRT. If non-monotone effects are also of interest, the Pearson and MIN2
test statistics are found to perform well, also for over- and under dominant
models \citep{JooRobustWellcome2009}. 

We have shown
how exact conditional enumeration is a valid and powerful competitor
to simulated permutations and asymptotic approximation for producing
$p$-values.
Drawing simulated permutations is an inefficient way of calculating an exact
conditional enumeration $p$-value for contingency tables of the size
and order used in genetic association studies. In our simulation study
calculating exact conditional
enumeration $p$-values was done in a fraction of a second, even for the largest sample
size considered in this presentation, $(5000, 15000)$.
Exact conditional enumeration should therefore
be preferred to simulated permutations. 

Further, it should be well known that $p$-values based on asymptotic
approximations need not be valid, especially for small
significance levels and unbalanced sample sizes, even when expected
cell counts are at least 5. In an extensive
simulation study we have seen that for the asymptotic approximation
for the test statistics studied here, the violation of test size may be
as large as 20 times the nominal level. Lastly, the fact that exact conditional enumeration methods give low
power compared to asymptotic methods for small sample sizes for 
$2\times 2$ contingency tables due to discreteness \citep{Lehmann1993}, is not transferrable to $2\times3$ contingency tables and the test statistics and sample sizes
under study. 
In conclusion we find that exact conditional enumeration
should also be preferred to asymptotic approximation,
both with respect to test size, power and computational considerations.

\section*{Software}
A \CC\ program using the GNU Scientific Library that takes the
entries of a $2\times
3$ table as input and gives the value of all seven test statistics
together with the asymptotic and exact conditional enumeration
$p$-values is available upon request.

\bigskip
\noindent\emph{Conflict of Interest:} None declared.

\bibliographystyle{chicago}
\bibliography{large}
\end{document}